\documentclass[12pt,a4paper]{article}
\usepackage[T2A]{fontenc}
\usepackage[utf8]{inputenc}
\usepackage[english]{babel}
\usepackage{amssymb}
\usepackage{amsmath}
\usepackage{graphicx}
\newcommand{\msun}{\,$M_{\odot}$}
\newcommand{\rsun}{\,$R_{\odot}$}
\newcommand{\ergs}{\,erg\,s$^{-1}$}
\newcommand{\kms}{\,km\,s$^{-1}$}

\begin{document}

\begin{center}
	\Large{\bf Type Ibn supernova SN~2010al: 
  Powerful mass loss half year prior to the explosion}	
	\vskip 5mm
	\small{N. N. Chugai}
	\vskip 5mm
	{\it Institute of Astronomy, Russian Academy of Sciences,
		Pyatnitskaya ul. 48, Moscow, 119017 Russia} \quad  \small{e-mail: nchugai@inasan.ru}\\
	
\end{center}
 
 \vskip 5mm
\centerline{\bf Abstract} 
\vskip 3mm
Type Ibn supernova SN~2010al is explored to infer 
  parameters of supernova and a circumstellar (CS) shell.
The CS interaction model combined with the spectral model of 4600\,\AA\ blend 
 suggests the explosion of a WR star with the energy of 
   $(1-1.5)\times10^{51}$\,erg inside a dense confined CS shell with the 
  mass of $\sim0.1$\msun\ 
  and kinetic energy of $\sim 10^{48}$\,erg.
The confined CS shell has been formed during the last 0.4 yr prior to the core collapse.

\newpage

\section{Introduction}
Supernova SN~2010al (type Ibn) is the core collapse supernova (CSSN) 
 associated with the explosion of a Wolf–Rayet (WR) star with 
  the signature of a dense  circumstellar (CS)  environment 
  (Pastorello et al. 2015).
The light curve of SNe~Ibn around the light maximum is powered by the ejecta  
 interaction with the dense CS gas (Moriya \& Maeda 2016) likewise  in the case     
  of SN 2006jc, another SN~Ibn (Chugai 2009).

The recent study of SNe~Ibn light curve sample including SN~2010al 
  (Maeda \& Moria (2022) led authors to conclude that the fast luminosity  
  decline after the light maximum reflects a steep CS density drop  
  $\rho \propto r^{-\omega}$ with $\omega \sim 3$ for $r > 10^{15}$\,cm and 
  the low $^{56}$Ni mass in supernova ejecta.
The steep CS density gradient in turn implies that the 
  mass loss rate increases as preSN approaches the  explosion (Maeda \& Moria 2022).
The presence of confined CS shell with the boundary radius of 
 $\sim 10^{15}$\,cm has been found earlier in type IIL SN~1998S (Chugai 2001) 
  and SNe~IIP, e.g. SN~2013fs (Yaron et al. 2017). 
These facts imply that some universal process in the core of massive stars gives 
  rise to a heavy mass loss year-decade prior to the core collapse.

The lack of clarity in understanding the origin of the vigorous mass loss 
  shortly before the CCSN explosion, and the 
  indication of the confined CS shell in SNe~Ib  motivate us 
 to explore the well observed SN~2010al to probe 
  parameters of the confined CS shell and the supernova envelope.
Among the appropriate tools for this task is the CS interaction model 
 (cf. Chugai 2001). 
Note that this reqires the description of  
  both the light curve and the expansion velocity; the latter is
  omitted in the recent model of SN~2010al. 
Moreover, the first spectrum of SN~2010al with the emission blend of He\,II\,4686\,\AA, N\,III\,4634, \,4641\,\AA\ 
(Pastorello et al. 2015) could provide us with an additional observational constraint for model parameters.
The point is that lines show narrow core and broad wings that is a signature of  
  the line emission and Thomson scattering in an opaque CS shell (Chugai 2001). 
The line profile modeling could permit us to recover the Thomson optical depth of 
 the CS shell and thus to validate the model of the CS shell.

I start with an overview of the CS interaction model including an extension on 
  the case of the adiabatic forward shock in the CS shell with 
  a steep density decline $\omega > 3$. 
I then describe the model aimed at the description of the 4600\,\AA\ blend.
Thereafter modeling results are presented with the discussion of 
 implications.

This study is based on SN~2010al spectrum (Pastorello et al. 2015) retrieved from the 
  WISeREP data base (Yaron \& Gal-Yam 2012).

\section{CS interaction model}

\subsection{Thin shell approximation}
\label{sec:tshell}

The hydrodynamics of the CS interaction will be described based on the thin shell 
  approximation that treats the swept up mass 
   between the reverse and forward shock as a thin shell driven by the ejecta  
    dynamical pressure (Giuliani 1981, Chevalier 1982, Chugai 2001).
In the relevant conditions a bulk of the thin shell is cold ($\sim 10^4$\,K) 
  and this shell can be dubbed the "cold dense shell" (CDS). 

The kinetic luminosity
  of the forward shock $L_{k,f}$ and reverse shock $L_{k,r}$ is converted into X-rays, which 
  are partially absorbed by the unshocked ejecta, CDS, and CSM thus giving rise 
    to the observed optical luminosity.
The X-ray luminosity at the age $t$ for a certain shock, e.g.,  forward shock, 
  is calculated as $L_{X,f} = \eta_f L_{k,f}$ with the radiation efficiency 
   $\eta_f = t/(t+t_{c,f})$, where $t_{c,f}$ is the cooling time of the postshock gas 
  in the forward shock.
The cooling time is calculated assuming $T_e = T_i$, the 
   shock density of $4\times$(preshock density), and the cooling function for 
  the hydrogen abundance X = 0.2 typical of WN stars (Hamann et al. 1991).
The fraction of X-rays from the forward shock of the radius $r_f$ that is  
 intercepted  by the unshocked ejecta and the CDS of the radius $r_{cds}$ is equal  
 to the dilution factor  $W = 0.5[1 - (1 - (r_{cds}/r_f)^2)^{1/2}]$.
The absorbed fraction of X-rays is calculated assuming thermal bremsstrahlung 
  spectrum for the shock temperature and absorption coefficient 
  $k_X = 100(E/1keV)^{8/3}$\,cm$^2$\,g$^{-1}$.

The model bolometric luminosity at the age $t$ suggests the instant re-emission of the 
   absorbed X-rays, provided the diffusion time for the CSM $t_{dif}(t) < t$,; otherwise 
  the luminosity is assumed to be $10^{40}$\ergs following the preSN luminosity of 
  SN~2020tlf with the enhanced pre-explosion mass loss (Jacobson-Galan et al. 2022).
The typical age $t$, when $t_{dif}(t) = t$ is $\sim 2$\,d. 

The CS density is set by the power law $\rho(r) = Ar^{-\omega}$ with $\omega < 3$ 
   for $r< R_k$ and $\omega > 3$ for $r > R_k$. 
A possible clumpiness of the CSM is ignored. 
The SN ejecta is set as a homologously expanding envelope ($v = r/t$) 
 with the density distribution $\rho(v) = \rho_0/[1 + (v/v_0)^8]$. 
Parameters $\rho_0$ and $v_0$ are specified via the ejecta mass $M$ and kinetic  
  energy $E$.

%============================================================
\begin{table}[t]
	\vspace{6mm}
	\centering
	{{\bf Table 1} Parameters of the CS interaction model}
	\label{tab:inter} 
	
	\vspace{5mm}\begin{tabular}{l|c|c|c|c|c} 
		\hline
 $M$~($M_{\odot}$) & $E$~($10^{51}$\,erg) &  
  $\omega_{in}$/$\omega_{out}$ & $R_k$ ($10^{15}$\,cm) & $M_{cs}$~($M_{\odot}$) & 
   $\tau(10d)$ \\
\hline
  5    &    1  & 1/4.9 & 1.4   &   0.14  & 3.4  \\

   \hline
	\end{tabular}
\end{table}
%===================================================================

The radiation output of the ejecta/CSM interaction is 
  determined by the kinetic energy  of ejecta external layers, which for the power law 
  SN density  distribution is the same for an infinite properly adjusted  combinations of $E$ and $M$. 
Particularly, for the ejecta density distribution $\rho \propto 1/v^n$ the effect of the CS interaction will be 
  the invariant provided $M$ and $E$ obey the relation $E \propto M^{(n-5)/(n-3)}$, 
  which reduces parameter degeneracy to the single parameter, e.g., $M$.
Adopting some value of the ejecta mass we are able to recover 
  the ejecta kinetic energy and CS density distribution based on the 
  light curve and the CDS velocity. 
With another choice of $M$ we immediately find the corresponding $E$ using 
  the above relation.  
For the fiducial model we consider the case of 5\msun\ ejecta, which corresponds to 
  the helium core of 6.5\msun\ for the main sequence 
   star of 21\msun\ (Woosley et al. 2002).

%====================================================================
\begin{figure}
\centering
\includegraphics[trim= 20 100 0 70,width=0.97\columnwidth]{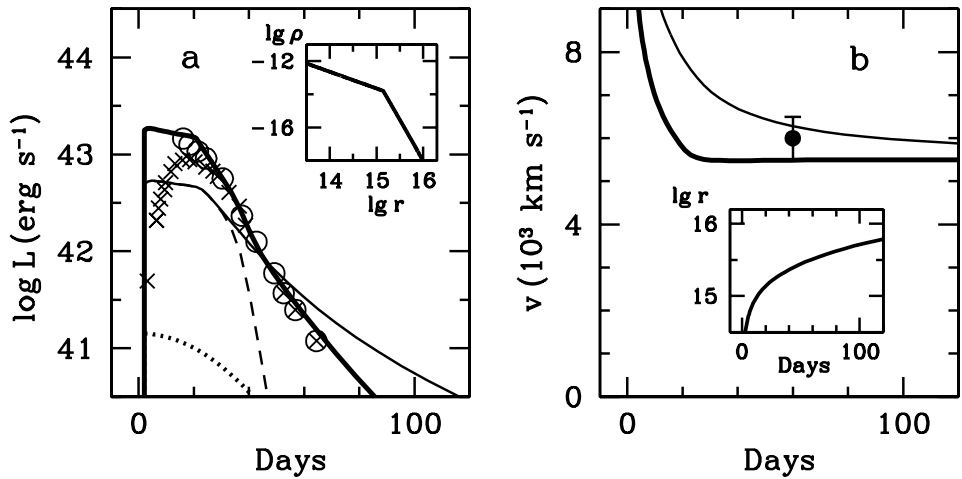}
\caption{
{\em Left panel}: Model bolometric light curve ({\em thick solid} line) overplotted 
  on two different versions of the observational data: the pseudo-bolometric light curves 
  inferred from the broad-band fluxes in the optical and near infrared domains ({\em crosses}) 
  and the light curve obtained including the near ultraviolet contribution ({\em circles}). 
The {\em thin} line is the bolometric luminosity powered by the forward shock in the thin   
 shell approximation, whereas the {\em dashed} line is the latter luminosity multiplied by  
  the guillotine factor.
The luminosity without CS interaction assuming preSN radius of 10\rsun\ 
  is shown by {\em dotted} line.
{\em Inset} shows the CS density distribution.
{\em Right panel}: The model CDS velocity ({\em thick} line) and the boundary velocity 
  of the unshocked ejecta ({\em thin} line). The maximal velocity recovered 
  from Ca\,II IR 
  triplet and He\,I 10830\,\AA\ line at about at the age of 60 days is shown by the 
  {\em  circle}. 
{\em Inset} shows the model CDS radius.
}
\label{fig:m5}
\end{figure}
%==================================================================

\subsubsection{Forward shock for $\omega > 3$}

Simulations reveal that to describe the steep luminosity decline of SN~2010al 
   after $t > t_{cr} \sim 40$\,d the thin shell model requires 
  a steep CS density gradient $(\omega > 5)$ in the outer zone $r > 10^{15}$\,cm.
In the case of strongly radiative forward shock the thin shell model is able to
  cope with this situation. 
However, if the forward shock becomes adiabatic 
  the thin shell model is not applicable anymore, since for $\omega > 3$ the
  adiabatic forward shock accelerates (Sedov 1993), whereas the CDS does not. 
 
In order to treat the adiabatic forward shock in the case $\omega > 3$ we apply a hybrid model.
Specifically, the reverse shock and the CDS expansion are treated using the thin shell model, 
   whereas the forward shock is described by the self-similar Sedov solution for the blast wave
    in a  non-uniform medium $\rho = Ar^{-\omega}$.
In this approach the shock radius is $r = Bt^{2/(5-\omega)}$ (Sedov 1993), 
  where $B$ depends on the blast wave energy $E$, CS density parameter $A$, and 
  adiabatic index.
However, we fix $B$ via matching the luminosity of the thin shell model with the 
  luminosity of the  detached adiabatic shock at $t = t_{cr} \sim 40$\,d, when the forward shock 
  enters the adiabatic regime.

The X-ray luminosity of the accelerating forward shock is estimated as 
 follows.
In the case $\omega > 3$ the swept up mass (total number of particles $N$) remains 
 almost constant since the most of the CS mass already swept up soon after 
 formation of the accelerated forward shock.
This means that the average postshock electron temperature 
 $T_e \propto E/N \sim$ const and thus the 
  cooling function $\Lambda(T_e)$ remains constant as well. 
Therefore, the X-ray luminosity of the forward shock 
 $L_{X,f} \propto r^{-3}N^2\Lambda \propto r^{-3} \propto t^{-6/(5-\omega)}$, while 
  the power absorbed by the CDS and unshocked ejecta is $L_f \propto WL_{X,f}$.
This is the maximal bolometric luminosity attributed to the forward shock.

For $r_f/r_{cds} \gg 1$ one gets  $W = (1/4)(r_{cds}/r_f)^2$.
With almost constant CDS velocity at the late stage one gets $W \propto t^{(6-2\omega)/(5-\omega)}$, 
so asymptotically $L_f \propto t^{-2\omega/(5-\omega)}$.
E.g, in the case $\omega = 4.5$ the luminosity is $L_f \propto t^{-18}$,
  a  steep decline with the negligible contribution of the forward shock 
  to the bolometric luminosity at the late time.
This behavior  can be described via the guillotine factor 
  $g = 1$ at $t < t_{cr}$ and zero otherwise.
The bolometric luminosity related to the accelerating forward shock 
 is then obtained by the multiplication of $g$ and the luminosity related to forward shock 
 of the thin shell model.
It is reasonable to use smooth version of the factor $g$   
\begin{equation}
g(t) = 1/[1 + (t/t_{cr})^s]\,,
\end{equation}
  where we adopt $s \sim 15$ and $t_{cr}$ is the moment when  
  cooling time $t_c$ meets the condition $t_c/t_{cr} = 0.5$.

\subsection{Modelling 4600\,\AA\ emission}

The 4600\,\AA\ emission blend in the first spectrum at about 10 days after the explosion 
  is composed by the He\,II\,4686\,\AA,  N\,III\,4634,\,4641\,\AA, and possibly 
  C\,III\,4647,\,4650\,\AA\ (Pastorello et al. 2015). 
We model the blend as a linear superposition of lines with the same 
  normalized profile. 
The spectrum of a single line is calculated using the Monte Carlo technique.
The model suggests that photon are emitted and scattered on electrons in the shell 
 with the inner radius $r_1$ coinciding with the CDS and the outer radius 
 $r_2 = 2.5r_1$.
The photosphere coincides with the CDS and is able to diffusively reflect photons 
 with the albedo $\Omega = 0.5$.
The density distribution corresponds to $\omega = 1$ in the inner zone of the CS interaction model,  $r < R_k$.
We adopt $n_e \propto \rho$, and emissivity $j \propto \rho^2$.
The electron temperature in the shell is assumed to be constant $T_e = 25000$\,K.

The CSM velocity recovered from absorption minima of narrow lines on days 12, 16, and 
  26\,d is 1000-1100\kms, 1050-1150\kms, and 1300-1400\kms\ respectively 
  (Pastorello et al. 2015).
The systematic velocity increase with time indicates that the velocity increases   
  along the radius. 
We set the radial dependence of CSM velocity by the linear relation
\begin{equation}
 u = (u_2 - u_1)(r - r_1)/(r_2 - r_1) + u_1\,,
 \end{equation}
   where $u_1$ is the CS gas velocity at the radius $r_1$ 
   and $u_2$ is the velocity of the undisturbed wind at $r_2$. 
 
The Thomson scattering takes into account Doppler shift between subsequent scatterings due to the expansion
 and the frequency redistribution 
 in the comoving frame caused by the electron thermal motion.
The latter is treated assuming angle-averaged 
  frequency redistribution function for the Thomson scattering on thermal electrons(Hummer \& Mihalas 1967).

%============================================================
\begin{table}[t]
	\vspace{6mm}
	\centering
	{{\bf Table 2} Model parameters for the 4600\,\AA\ blend}
	\label{tab:blend} 
	
	\vspace{5mm}\begin{tabular}{l|c|c|c|c} 
		\hline
Model & $\tau$ &   C/N   & $u_1$ (km/s) & $u_2$ (km/s) \\
\hline
A      &   3.5    &  0.19   & 400   & 1300 \\   
B      &   1      &  0.19   & 400   & 1300 \\ 
C      &   3.5    &   0     & 400   & 1300 \\ 
D      &   3.5    &  0.19   & 1000  & 1000 \\ 
   \hline
	\end{tabular}
\end{table}
%===================================================================

\section{Parameters of supernova and CS shell}

The bolometric light curve and the expansion velocity are described by the optimal 
 model (Figure \ref{fig:m5}) with parameters presented in Table~1.
The Table includes ejecta mass, ejecta energy, power law index 
 of the CS density in inner ($r < R_k$) and outer zones, $R_k$ value, 
 the CS shell mass in the range $r \leq R_k$, 
and the Thomson optical depth of the CS shell outside the CDS on day 10.
At the stage $t \lesssim 40$\,d the luminosity related to the reverse and forward shocks 
  are comparable, whereas at the later stage the luminosity is determined 
  entirely by the reverse shock.  
Remarkably, the model velocity of the CDS and boundary velocity of unshocked ejecta 
  are consistent with the maximal expansion velocity estimated from He\,I\,10830\,\AA\ 
  and calcium triplet Ca\,II 8600\,\AA\ in the spectra on day 60.

The light curve in combination with the CDS velocity  
  permit us to find the explosion energy for the adopted ejecta mass of 5\msun. 
It is already emphasized that for the outer ejecta power law density $\rho \propto 1/v^n$ 
the energy should scale as $E \propto M^{(n-5)/(n-3)}$.
Particularly, for $n = 8$ and twice as high ejecta mass the 
  energy must be by 1.516 times larger, i.e., M = 10\msun\ ejecta with 
  the energy $E = 1.52\times10^{51}$\,erg produces the 
 same result as the model with $M = 5$\msun, the fact we also confirmed numerically.
The 10\msun\ ejecta corresponds to 11.5\msun\ preSN or $\approx 40$\msun\ 
  main sequence progenitor  (Woosley et al 2002).
A successful explosion of CCSNe with the formation of neutron star 
  occurs only for stars with the initial mass 
 $< 40$\msun, while the mass range $\lesssim 25$\msun\ 
 gives rise to SNe~IIP (Heger et al. 2003). We conclude therefore 
 that SNe~Ibn progenitors 
  originate from the mass range $25 \lesssim M < 40$\msun. 
This means that the explosion energy of SN~2010al 
  is in the range of $(1-1.5)\times10^{51}$\,erg. 
 
In the context of the progenitor mass of a high interest is the 
 $^{56}$Ni mass in SN~2010al ejecta.
Based on the CS interaction model we find that $M_{Ni} < 0.01$\msun. 
The model independent estimate follows from the late observational bolometric 
  luminosity, $M_{Ni} \leq 0.015$\msun, which is consistent with the upper limit 
  found earlier $M_{Ni} < 0.02$\msun\ (Maeda \& Moriya 2022).

%====================================================================
\begin{figure}
\includegraphics[trim= 40 120 0 0,width=0.9\columnwidth]{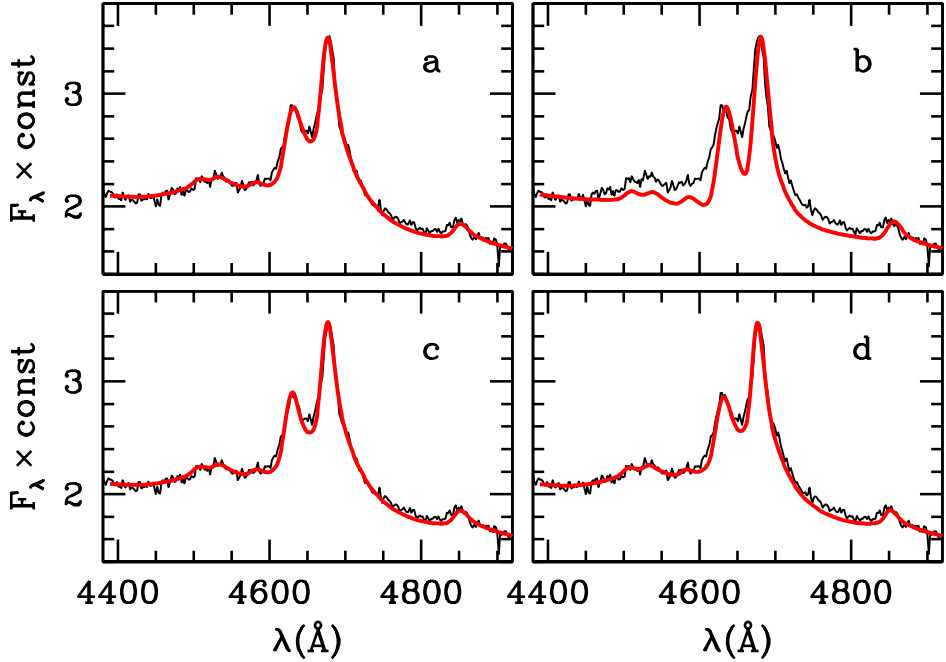}
\caption{
Model spectrum of the blend composed by He\,II, N\,III and possible C\,III lines 
{\em red} line overplotted on the observed spectrum. Panels a, b, c, d show
 models A, B, C, D, respectively (Table 2). 
}
\label{fig:blend}
\end{figure}
%==================================================================

The CS interaction model is supported by the 4600\,\AA\ blend modeling 
 (model A, Figure \ref{fig:blend}, Table 2).
Apart from He\,II\,4686\,\AA,  N\,III\,4637\,\AA, the model includes N\,III 
 4515, 4544, and 4592\,\AA\ lines, C\,III emission, and H$\beta$.
The Table 2 includes the CS shell optical depth, C/N that stands for the 
 flux ratio C\,III\,4648/N\,III\,4637, the CS velocity at the radii $r_1$ and 
 $r_2$.
The CS density distribution ($\omega = 1$) and the optical depth of the CS shell   
  outside the CDS ($\tau = 3.4$) areconsistent with  parameters of the CS 
   interaction model. 

The line profile is weakly sensitive to the 
  electron temperature variation in the range 20000-30000\,K; we adopt 
 $T_e = 2500$\,K.
By 3.6 days later the black body temperature is 21000\,K (De la Rosa et al 2016), 
 which is in line with the adopted electron temperature at the earlier age.  
Photospheric albedo ($\Omega$) also does not significantly affect the profile either; we adopt $\Omega = 0.5$.
The model B (Figure \ref{fig:blend}, Table 2)
  shows the pronounced effect of the low Thomson optical depth ($\tau = 1$).
The model C without C\,III line suggests the lack of strong evidence for 
  the presence of C\,III line, although the fit at about 4650\,\AA\ is somewhat 
 worse.
The model D with the constant CS expansion velocity of 1000\kms\ 
    fits the red wing of the He\,II line noticeably worse compared to the model A.

%============================================================
\begin{table}[t]
	\vspace{6mm}
	\centering
	{{\bf Table 3} Confined CS shell in CCSNe}
	\label{tab:inter} 
	
	\vspace{5mm}\begin{tabular}{l|c|c|c|c|c|c} 
		\hline
SN type  &   SN      &  $M_{cs}$~($M_{\odot}$) & $u_{cs}$ (km/s) &  $E_{cs}$ (erg) & 
 $^{56}$Ni ($M_{\odot}$) & $t_{cs}$ (yr)\\ 
\hline
  &            &                    &        &   &    \\

 SN~IIP    &  2013fs    &   0.003$^a$     & 50    &  $7\times10^{43}$ & 0.05$^d$ & $\sim10$ \\
   &            &                    &       &    &   \\
 SN~IIL   &  1998S     &     0.1$^b$     &   40      &   $2\times10^{45}$ & 0.15$^e$ & $\sim10$\\
          &            &                    &        &   &   \\
 SN~Ibn         &  2010al    &   0.14$^c$    &  $10^3$    &  $10^{48}$ & $<0.015^c$ & 0.4\\ 
   \hline
\end{tabular}
\parbox[]{12cm}{\small $^a$\,Yaron et al (2017), $^b$\,Chugai (2001), $^c$\,this paper, 
  $^d$\,Chugai (2020), \\ $^e$\,Fassia et al (2000)}
\end{table}
%===================================================================

\section{Discussion and conclusions}

The goal of this paper has been to explore the well observed type Ibn supernova 
 SN~2010al in order to infer parameters of the supernova and the confined CS shell.
The CS interaction model and the model of 4600\,\AA\ emission provide 
  us with a picture of the WR progenitor explosion with the energy of 
   $(1-1.5)\times10^{51}$\,erg inside a confined CS shell ($\sim 10^{15}$\,cm) 
  with the mass of 0.14\msun.
Remarkably, the explosion energy range is well within the neutrino-driven explosion 
  $E \lesssim 2\times10^{51}$ (Janka 2017).
  
The confined CS shell with the mass of $M_{cs} = 0.14$\msun\ and the expansion velocity of 
  $u \approx 1100$\kms\ within the radius $R_k = 1.4\times10^{15}$\,cm  
  is therefore produced during the last $t_{cs} = R_k/u \sim 0.4$\,yr
   due to tremendous mass loss rate $M_{cs}/t_{cs} \sim 0.3$\msun\,yr$^{-1}$. 
The overall energy of 
  this event is $E_{cs} = (1/2)M_{cs}u^2 \sim 1.7\times10^{48}$\,erg and the average kinetic luminosity of 
  the mass loss thus is $E_{cs}/t_{cs} \sim 10^{41}$\ergs. 

SN~2010al shows the maximal energy of the confined CS shell among core-collapse 
 SNe with similar CS shell (here we do not include events similar to SN~1994W 
  and SN~2006gy).
Table 3 presents parameters of three well studied  CCSNe of different types with 
 confined CS shell.
The Table includes the mass of the confined CS shell, velocity of the CS gas, 
  kinetic energy of the CS shell, $^{56}$Ni mass in the supernova ejecta, and the  
 duration of the heavy mass loss responsible for the CS shell formation. 
These supernovae compose the sequence along the energy of CS shell: 
  SN~IIP $\rightarrow$ SN~IIL $\rightarrow$ SN~Ibn 
     with a large energy increment.
It is sensible to suggest that the order reflects the growing progenitor mass along the 
 sequence.  
In that case the central source responsible for the mass loss operates according to 
  the rule: the larger progenitor mass, the larger energy of hydrodynamic perturbations  
  is transferred from the core to outer layers.

The theory of massive star evolution predicts that the oxygen burning 
 in the core takes less time for progenitor with larger initial mass.
  For the 25\msun\ star the oxygen burning time is of 0.4 yr 
  (Woosley et al. 2002), comparable to the time for the CS shell formation in 
  SN~2010al. 
This indicates that the high energy of the CS shell of SN~2010al might be 
  related to the  main sequence star of $\sim 25$\msun.

Processes involved in the generation of hydrodynamic perturbations responsible 
  for the powerful mass loss of pre-collapse supernovae are far from clear.
An interesting possibility is that the vigorous core convection might  generate 
  powerful flux of acoustic waves (Quataert \& Shiode 2012). 
The  WR presupernova of SN~2010al with the high energy of the expelled shell 
 indicates that that a slow mass loss is highly unlikely.
A more appropriate mass loss regime is the shell ejection by a shock wave with the 
 energy of the order of $10^{48}$\,erg.

If the energy of perturbations responsible for the vigorous mass loss shortly before 
  the collapse increases with the progenitor mass , then the 
 small amount of $^{56}$Ni in SN~2010al ejecta could be related to the fallback 
   most of $^{56}$Ni onto the neutron star, which is the case for  
  massive progenitor (Woosley et al 2002).
Note that the fallback of significant amount of $^{56}$Ni-rich matter suggests formation of 
massive neutron star in SN~2010al. The existence of neutron stars with masses up to 2\msun\  
is the observational fact  (cf. Fonseca et al. 2001).
\vspace{0.5cm}

I thank Lev Yungelson for useful discussions. This research is  
supported by RFBR and DFG grant 21-52-12032.

\vspace{1cm}

\centerline{\bf References}
\vspace{0.6cm}
%==================================================================

\noindent Chevalier R.A., ApJ {\bf 259}, 302 (1982) \\
Chugai N.N. MNRAS, {\bf 494}, L86 (2020)\\
Chugai N.N., MNRAS, {\bf 400}, 866 (2009)\\
Chugai N.N., Blinnikov S.I., Fassia A. et al. MNRAS {\bf 330}, 473 (2002)\\
Chugai N.N., MNRAS, {\bf 326}, 1448 (2001)\\
De la Rosa1 J., Roming P., Pritchard T,, and Fryer C. ApJ {\bf 820}, 74 (2016) \\
Fassia A., Meikle W.P.S., Vacca W.D. et al. MNRAS {\bf 318}, 1093 (2000)\\
Fonseca E., Cromartie H.T., Pennucci T.T. et al. ApJ {\bf 915}, L12 (2021)\\
Giuliani J. L. ApJ {\bf 245}, 903 (1981)\\ 
Heger A., Fryer C.L., Woosley S.E. et al. Astrophys. J. {\bf 591}, 288 (2003)
Hummer D. G., Mihalas D., ApJ {\bf 150}, L57 (1967)\\
Janka H.-T., 2017, Neutrino-Driven Explosions. p. 1095\\
Jacobson-Galán, W. V.; Dessart, L.; Jones, D. O 2022ApJ...924...15J \\
Maeda K., Moriya T.  eprint arXiv:2201.00955 (2022) \\
Morya T., Maeda K.  ApJ {\bf 824}, 100 (2016)\\
Quataert E., Shiode J. MNRAS {\bf 423}, L92 (2012)\\
Pastorello A., Benetti S., Brown P.J. Tsvetkov D.Y. et al.,
 MNRAS {\bf 449}, 1921 (2015)\\
Sander A., Hamann W.-R. , and Todt H. Astron. Astrophys. {\bf 540}, 144 (2012)
Sedov L.I. {\it Similarity and dimensional methods in mechanics} 
  (CRC Press, NW Boca Raton, 1993) \\
Woosley S.E., Heger A., Weaver T.A., Rev. Mod. Phys. {\bf 74}, 1015 (2002)\\
Yaron O., Perley D.A., Gal-Yam A. et al. Nature Physics, {\bf 13}, 510 (2017)\\
Yaron O., Gal-Yam A., Publ. Astron. Soc. Pacific {\bf124}, 668 (2012) \\ 

%=====================================================

%========================================================

\end{document}